\documentclass[twocolumn,prb,preprintnumbers,amsmath,amssymb,floats,citeautoscript,nobalancelastpage,showpacs]{revtex4}

\usepackage{graphicx}
\usepackage{dcolumn}
\usepackage{bm}
\usepackage{times}

\begin{document}

\title{Ga$^{3+}$-substitution effects in the weak ferromagnetic oxide LaCo$_{0.8}$Rh$_{0.2}$O$_{3}$}

\author{Shinichiro Asai}
\author{Noriyasu Furuta}
\author{Ryuji Okazaki}
\author{Yukio Yasui$^{\ast}$}
\author{Ichiro Terasaki}

\affiliation{Department of Physics, Nagoya University, Nagoya 464-8602, Japan}

\begin{abstract}
Magnetization and x-ray diffraction have been measured on polycrystalline samples of LaCo$_{0.8-y}$Rh$_{0.2}$Ga$_{y}$O$_{3}$ for $0 \leq y \leq 0.15$ in order to understand the spin state of Co$^{3+}$ through the Ga$^{3+}$ substitution effect.
The ferromagnetic order in LaCo$_{0.8}$Rh$_{0.2}$O$_{3}$ below 15 K is dramatically suppressed by the Ga$^{3+}$ substitution, where the ferromagnetic volume fraction is linearly decreased.
The normal state susceptibility also systematically decreases with the Ga content, from which we find that one Ga$^{3+}$ ion reduces 4.6 $\mu_{\rm B}$ per formula unit.
We have evaluated how the concentration of the high-spin state Co$^{3+}$ changes with temperature by using an extended Curie-Weiss law, and have found that the substituted Rh$^{3+}$ ion stabilizes the high-spin state Co$^{3+}$ ion down to low temperatures.
We find that Ga$^{3+}$ preferentially replaces the high-spin state Co$^{3+}$, which makes a remarkable contrast to our previous finding that Rh$^{3+}$ preferentially replaces the low-spin state Co$^{3+}$.
These results strongly suggest that the magnetically excited state of LaCoO$_{3}$ at room temperature is a mixed state of high-spin state Co$^{3+}$ and low-spin state Co$^{3+}$.
\end{abstract}

\pacs{75.30.Wx, 75.47.Lx, 75.50.Dd}

\maketitle

\section{Introduction}
The perovskite cobalt oxide LaCoO$_{3}$ has been investigated for their various physical properties.
The five-fold degenerate 3\textit{d} orbitals in Co$^{3+}$ ions are split into the doubly-degenerate \textit{e}$_{g}$ orbitals in the upper energy level and the triply-degenerate \textit{t}$_{2g}$ orbitals in the lower energy level due to the Coulomb interaction between the Co$^{3+}$ ions and the O$^{2-}$ ions coordinated octahedrally.
The six \textit{d} electrons in 3\textit{d} orbitals of Co$^{3+}$ ions are fully occupied in \textit{t}$_{2g}$ orbitals in the low-spin state (\textit{t}$_{2g}^{6}$; S = 0), when the crystal field splitting is larger than the Hund coupling.
The Co$^{3+}$ ions take high-spin state (\textit{e}$_{g}^{2}$\textit{t}$_{2g}^{4}$; S = 2) for the opposite condition.
LaCoO$_{3}$ is a fascinating material with respect to the spin state because the Co$^{3+}$ ion in LaCoO$_{3}$ shows spin-state crossover with temperature~\cite{Physica.30.1600}, magnetic field~\cite{PRB.67.140401}, or pressure~\cite{JPSJ.78.093702}.
It is theoretically suggested that the Co$^{3+}$ ion in LaCoO$_{3}$ can take the intermediate-spin state (\textit{e}$_{g}^{1}$\textit{t}$_{2g}^{5}$; S = 1), when the hybridization between the \textit{e}$_{g}$ orbitals and O 2\textit{p} orbitals is taken into account.~\cite{PRB.54.5309}

The magnetic susceptibility of LaCoO$_{3}$ decreases with decreasing temperature below 90 K, takes a maximum at around 90 K, and has the Curie-Weiss like behavior around room temperature.~\cite{Physica.30.1600}
This indicates that the Co$^{3+}$ ion is non-magnetic in the ground state, and a magnetic state is thermally activated above 90 K.
The change from the non-magnetic ground state to the magnetically excited state is also observed in NMR~\cite{,JPSJ.64.3967} and neutron scattering~\cite{PRB.40.10982}.
The excited state of LaCoO$_{3}$ at room temperature has been theoretically and experimentally studied~\cite{Physica.30.1600,PRB.67.140401,JPSJ.78.093702,PRB.54.5309,PRB.40.10982,JPSJ.64.3967,PR.155.932,PRB.71.024418,PRL.97.176405,PRB.67.144424,PRB.66.094404,PRB.57.10705,PRB.55.4257,PRB.67.224423,PhysRevB.55.R8666,JPSJ.70.3296}, which is still controversial. 
In 1966, Raccah and Goodenough proposed an NaCl-type spin-state order consisting of the high-spin and low-spin states of Co$^{3+}$ as the excited state of LaCoO$_{3}$.~\cite{PR.155.932}
They suggested that the displacement of O$^{2-}$ ions toward the low-spin state Co$^{3+}$ stabilizes the excited state due to the different crystal fields.
No evidence for the spin-state order has been observed until now.
Aside from such a static order, a dynamically mixed state of the high-spin and low-spin states (HS-LS model) is suggested from soft x-ray absorption spectroscopy~\cite{PRL.97.176405}, heat capacity combined with the magnetic susceptibility~\cite{PRB.71.024418,PRB.67.144424}, ESR~\cite{PRB.66.094404}, and the unrestricted Hartree-Fock calculation~\cite{PRB.57.10705}.
In 1996, Korotin \textit{et al.} proposed another model that the excited state of LaCoO$_{3}$ is the intermediate-spin state (IS model)~\cite{PRB.54.5309}, which is qualitatively different from HS-LS model.
This model is supported by the pressure dependence of the structure~\cite{PRB.67.140401}, x-ray photoemission spectroscopy~\cite{PRB.55.4257}, neutron diffraction~\cite{JPSJ.70.3296}, synchrotron x-ray powder-diffraction~\cite{PRB.67.224423}, and infrared spectroscopy.~\cite{PhysRevB.55.R8666}

Kyomen \textit{et al.}~\cite{PRB.67.144424} have found that the substitution effects of two non-magnetic ions, Ga$^{3+}$ (3$d^{10}$; S = 0) and Rh$^{3+}$ ($t_{2g}^{6}$; S = 0), for Co$^{3+}$ differ from each other.
The magnetic susceptibility of LaCo$_{1-x}$Ga$_{x}$O$_{3}$ decreases with increasing Ga content, which indicates that the Ga$^{3+}$ substitution for Co$^{3+}$ does not change the non-magnetic ground state. 
On the other hand, a Curie-Weiss like susceptibility develops at low temperature for $x > 0.04$ in LaCo$_{1-x}$Rh$_{x}$O$_{3}$.
In contrast to the difference of the substitution effects for the magnetization between two species of non-magnetic ions, the electrical resistivity and Seebeck coefficients of LaCo$_{1-x}$Ga$_{x}$O$_{3}$~\cite{IM451026} and LaCo$_{1-x}$Rh$_{x}$O$_{3}$~\cite{JSSC.183.1388} at high temperature show similar behavior.
GGA+\textit{U} calculation indicates that the Rh$^{3+}$ substitution stabilizes the high-spin state Co$^{3+}$ due to elastic interaction between the cations with different ionic radius and electronic interaction associated with unfilled 4\textit{d} shell of Rh$^{3+}$.~\cite{PhysRevB.85.134401} 
In our previous study, we measured the x-ray diffraction and magnetization of the polycrystalline samples of LaCo$_{1-x}$Rh$_{x}$O$_{3}$ in the range of $0 \leq x \leq 0.9$ in order to investigate the magnetism induced in the solid solution of the two non-magnetic end phases of LaCoO$_{3}$ and LaRhO$_{3}$,~\cite{JPSJ.80.104705} and found a ferromagnetic ordering below 15 K in the range of $0.1 \leq x \leq 0.4$.
This ferromagnetic ordering is driven only by Co$^{3+}$ ions, and is distinguished from that observed in La$_{1-x}$Sr$_{x}$CoO$_{3}$~\cite{Physica.19.120} due to the double exchange interaction between the low-spin state Co$^{4+}$ and the intermediate-spin state Co$^{3+}$.
We further found that the effective magnetic moment of LaCo$_{1-x}$Rh$_{x}$O$_{3}$ evaluated at room temperature is independent of Rh content $x$ for $0 \leq x \leq 0.5$.
It is also found that the lattice volume for LaCo$_{1-x}$Rh$_{x}$O$_{3}$ is larger than that expected from Vegard's law.
These results suggest that a Rh$^{3+}$ ion is preferentially substituted for a low-spin state Co$^{3+}$ ion and that the excited state of LaCoO$_{3}$ is described with HS-LS model.

In this paper, we have measured the x-ray diffraction and the magnetization of LaCo$_{0.8-y}$Rh$_{0.2}$Ga$_{y}$O$_{3}$ ($0 \leq y \leq 0.15$) in order to clarify the Ga$^{3+}$ substitution effect on the ferromagnetism as well as the spin state of Co$^{3+}$.
The reason why LaCo$_{0.8}$Rh$_{0.2}$O$_{3}$ is chosen as the initial phase for the Ga$^{3+}$ substitution is that this composition gives the largest magnetization at low temperature with the highest Curie temperature.
\section{Experiments}
Polycrystalline samples of LaCo$_{0.8-y}$Rh$_{0.2}$Ga$_{y}$O$_{3}$ ($0 \leq y \leq 0.15$) were prepared by a conventional solid-state reaction.
A mixture of La$_{2}$O$_{3}$ (3N), Co$_{3}$O$_{4}$ (3N), Rh$_{2}$O$_{3}$ (3N), and Ga$_{2}$O$_{3}$ (4N) with stoichiometric molar ratios was ground and calcined for 24 hour at 1000 $^{\circ}$C in air.
The calcined powder was ground, pressed into a pellet, and sintered for 48 hour at 1200 $^{\circ}$C in air.

X-ray diffraction was measured with a Rigaku RAD-I\hspace{-.1em}IC (Cu K$\alpha$ radiation), and no impurity phases were detected for the prepared samples.
Magnetization in field cooling (FC) and zero field cooling (ZFC) processes were measured using a superconducting quantum interference device (SQUID) magnetometer (Quantum Design MPMS) from 5 to 300 K in an applied field of 1 T for LaCo$_{0.8-y}$Rh$_{0.2}$Ga$_{y}$O$_{3}$ ($0 \leq y \leq 0.15$).
Magnetization-field (\textit{M} - \textit{H}) curves at \textit{T} = 2 K for LaCo$_{0.8-y}$Rh$_{0.2}$Ga$_{y}$O$_{3}$ ($0 \leq y \leq 0.15$) were measured in sweeping $\mu_{0}$\textit{H} from 0 to 7 T, and from 7 to 0 T.
\section{Results and discussion}
From the x-ray diffraction measurements, we find that the crystal system of the samples at room temperature is orthorhombic, which means that Ga$^{3+}$ substitution does not change the orthorhombic structure of LaCo$_{0.8}$Rh$_{0.2}$O$_{3}$.
\begin{figure}[!htb]
\begin{center}
\includegraphics[width=70.0mm,clip]{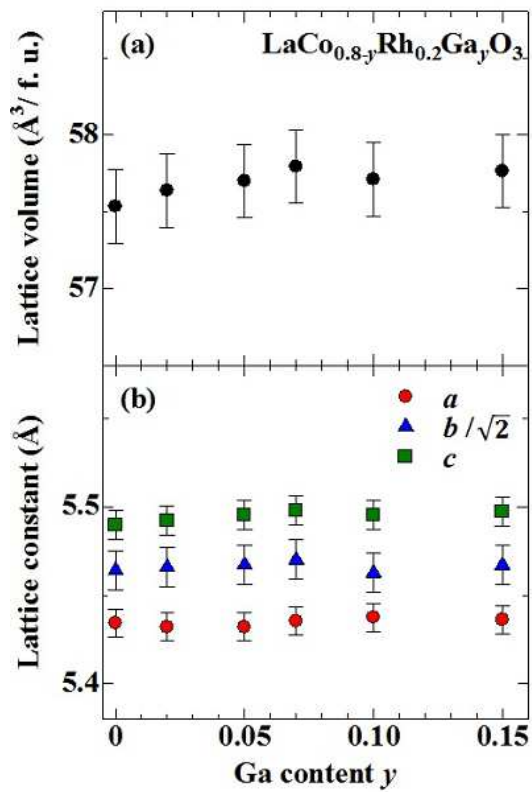}\\
\caption{(Color online) The Ga content dependence of (a) the lattice volume and (b) the lattice constants for LaCo$_{0.8-y}$Rh$_{0.2}$Ga$_{y}$O$_{3}$ ($0 \leq y \leq 0.15$).}
\end{center}
\end{figure}
Figures 1(a) and 1(b) show the lattice volume per formula unit and the lattice constants for LaCo$_{0.8-y}$Rh$_{0.2}$Ga$_{y}$O$_{3}$ ($0 \leq y \leq 0.15$), respectively.
The two quantities are essentially independent of Ga content, which remarkably differs from the Rh$^{3+}$ substitution effect, where the lattice volume increases by 1 \AA $^{3}$ from LaCo$_{0.8}$Rh$_{0.2}$O$_{3}$ to LaCo$_{0.65}$Rh$_{0.35}$O$_{3}$.~\cite{JPSJ.80.104705}

\begin{figure}[!htb]
\begin{center}
\includegraphics[width=70.0mm,clip]{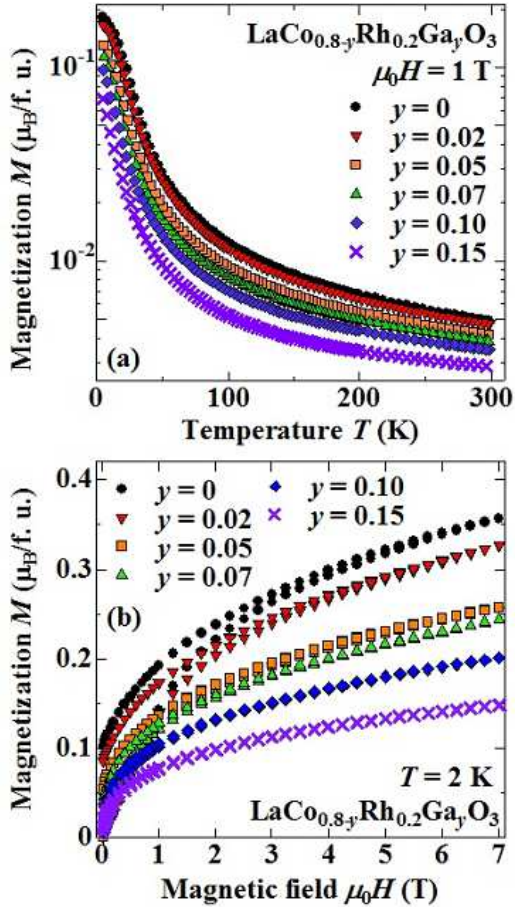}\\
\caption{(Color online) (a) Temperature dependence of the magnetization for LaCo$_{0.8-y}$Rh$_{0.2}$Ga$_{y}$O$_{3}$ ($0 \leq y \leq 0.15$) taken in 1 T.
(b) Magnetization - field curves for LaCo$_{0.8-y}$Rh$_{0.2}$Ga$_{y}$O$_{3}$ ($0 \leq y \leq 0.15$) at 2 K.}
\end{center}
\end{figure}
Figure 2(a) shows the temperature dependence of the magnetization for LaCo$_{0.8-y}$Rh$_{0.2}$Ga$_{y}$O$_{3}$ per formula unit in the range of $0 \leq y \leq 0.15$ in 1 T.
Magnetization in all the temperature range decreases with increasing Ga content, whereas the Curie-Weiss-like temperature dependence remains intact.
Figure 2(b) shows the applied field dependence of the magnetization (\textit{M} - \textit{H} curve) at $T = 2$ K for LaCo$_{0.8-y}$Rh$_{0.2}$Ga$_{y}$O$_{3}$ ($0 \leq y \leq 0.15$).
The spontaneous magnetization observed for $y = 0$ is strongly suppressed by Ga$^{3+}$ substitution, and is not visible for $y = 0.15$.
It indicates that the weak ferromagnetism is severely suppressed by Ga$^{3+}$ substitution.

Here we compare the effects of the Ga$^{3+}$ substitution on the magnetization with those of the Rh$^{3+}$ substitution.
\begin{figure}[!htb]
\begin{center}
\includegraphics[width=85.0mm,clip]{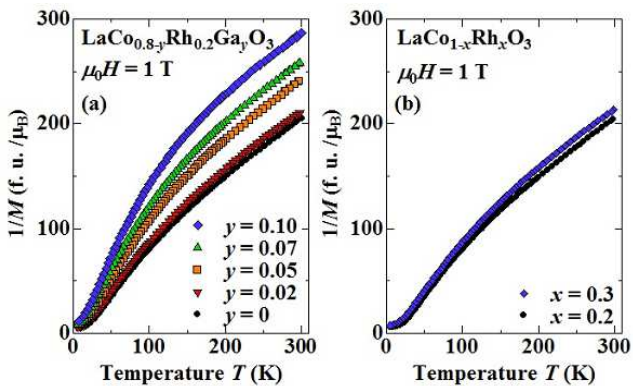}\\
\caption{(Color online) Temperature dependence of the reciprocal magnetization for (a) LaCo$_{0.8-y}$Rh$_{0.2}$Ga$_{y}$O$_{3}$ ($0 \leq y \leq 0.1$) and (b) LaCo$_{1-x}$Rh$_{x}$O$_{3}$ ($x$ = 0.2 and 0.3).}
\end{center}
\end{figure}
Figures 3(a) and 3(b) show the temperature dependence of the reciprocal magnetization for LaCo$_{0.8-y}$Rh$_{0.2}$Ga$_{y}$O$_{3}$ ($0 \leq y \leq 0.1$) and LaCo$_{1-x}$Rh$_{x}$O$_{3}$ ($x$ = 0.2 and 0.3), respectively.
The inverse magnetization increases by the Ga$^{3+}$ substitution much more drastically than by the Rh$^{3+}$ substitution.
Considering that $y = 0.1$ and $x = 0.1$ correspond to 10\% substitution for LaCo$_{0.8}$Rh$_{0.2}$O$_{3}$, we should emphasize that at least two kinds of Co$^{3+}$ ion exist in LaCo$_{0.8}$Rh$_{0.2}$O$_{3}$; the one is non-magnetic, and the other is magnetic.
This favors the HS-LS model rather than the uniform IS model.

The HS-LS model is also supported by the Ga content dependence of the lattice volume and the lattice constants for LaCo$_{0.8-y}$Rh$_{0.2}$Ga$_{y}$O$_{3}$ shown in Fig. 1.
The experimental results naturally indicate that the Ga$^{3+}$ ion is substituted for the Co$^{3+}$ ion with the same ionic radius.
According to the literature~\cite{Shannon}, the ionic radius of the Ga$^{3+}$ ion (0.62 \AA) is larger than that of the low-spin state Co$^{3+}$ (0.54 \AA) but is close to that of the high-spin state Co$^{3+}$ (0.61 \AA).
It seems that Ga$^{3+}$ is preferentially substituted for the high-spin state Co$^{3+}$.

Let us further discuss the Ga$^{3+}$ substitution effects on the spin system.
We fit the \textit{H}/\textit{M} - \textit{T} curves from 40 to 70 K shown in Fig. 3(a) using the Curie-Weiss law given by
\begin{equation}
\frac{M}{\mu_{0} H}=\frac{N\mu_{\rm B}^{2}\mu^{2}_{\rm eff}}{3k_{\rm B}(T-\theta)},
\end{equation}
and have evaluated the effective magnetic moment $\mu_{\rm eff}$ and the Curie temperature $\theta$.
\begin{figure}[tb]
\begin{center}
\includegraphics[width=70.0mm,clip]{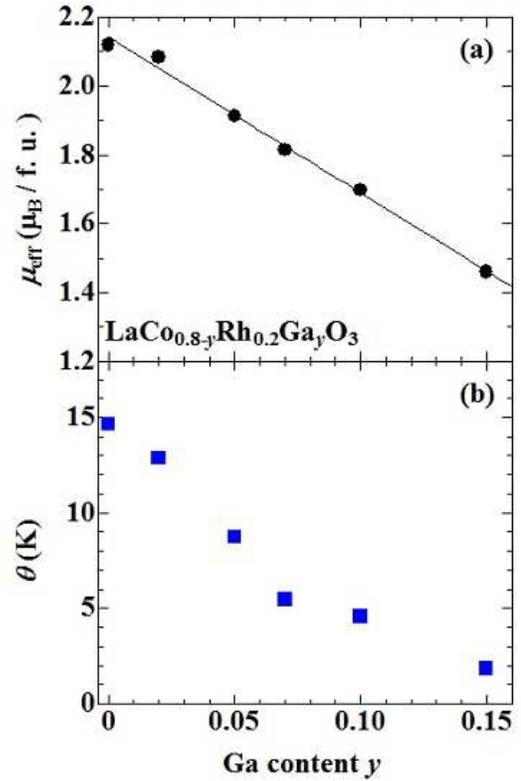}\\
\caption{(Color online) Ga content dependence of (a) the effective magnetic moment and (b) the Curie temperature for LaCo$_{0.8-y}$Rh$_{0.2}$Ga$_{y}$O$_{3}$ ($0 \leq y \leq 0.15$).
The solid line in Fig. 5(a) represents the linear fit to the data.}
\end{center}
\end{figure}
Figure 4(a) shows the Ga content dependence of $\mu_{\rm eff}$.
The evaluated $\mu_{\rm eff}$ is 2.1 $\mu_{\rm B}$ for $y = 0$, which linearly decreases with increasing Ga content.
We evaluate from the slope of the solid line shown in Fig. 4(a) that  one Ga$^{3+}$ ion decreases 4.6 $\mu_{\rm B}$, which is consistent with our claim that Ga$^{3+}$ replaces the high-spin state Co$^{3+}$ of 4.9 $\mu_{\rm B}$.
The Ga content dependence of $\theta$ is shown in Fig. 4(b), where $\theta$ linearly decreases with increasing Ga content.
This indicates that the magnetic interaction between the high-spin state Co$^{3+}$ ions is suppressed by Ga$^{3+}$ substitution, which further consolidates our claim that the high-spin state Co$^{3+}$ ions are substituted by Ga$^{3+}$ ions.

Next, we discuss the \textit{M} - \textit{H} curves of LaCo$_{0.8-y}$Rh$_{0.2}$Ga$_{y}$O$_{3}$ ($0 \leq y \leq 0.15$) shown in Fig. 2(b).
\begin{figure}[!htb]
\begin{center}
\includegraphics[width=70.0mm,clip]{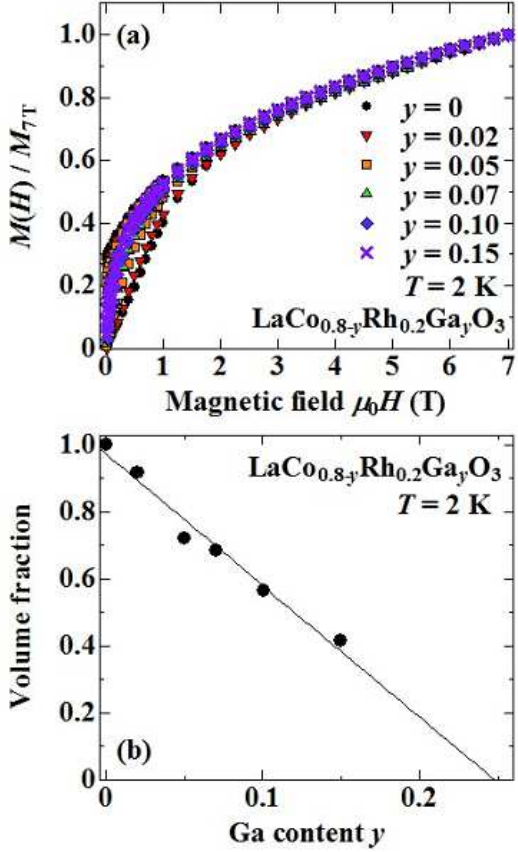}\\
\caption{(Color online) (a)Magnetization - field curves normalized by the value of the magnetization taken in 7 T for LaCo$_{0.8-y}$Rh$_{0.2}$Ga$_{y}$O$_{3}$ ($0 \leq y \leq 0.15$) at 2 K.
(b) Ga content dependence of the volume fraction (see text) for LaCo$_{0.8-y}$Rh$_{0.2}$Ga$_{y}$O$_{3}$ ($0 \leq y \leq 0.15$) at 2 K.
The solid line represents the linear fit of the data.}
\end{center}
\end{figure}
Figure 5(a) shows \textit{M} - \textit{H} curves normalized by the value of magnetization at $\mu_{0}$\textit{H} = 7 T for LaCo$_{0.8-y}$Rh$_{0.2}$Ga$_{y}$O$_{3}$ ($0 \leq y \leq 0.15$), where all the curves fall onto a single curve above 3 T.
This indicates that the Ga$^{3+}$ substitution dilutes the magnetization of LaCo$_{0.8}$Rh$_{0.2}$O$_{3}$.
The normalization factor is shown in Fig. 5(b), which represents the volume fraction of the magnetic phase.
The volume fraction linearly decreases with increasing Ga content, which indicates that the high-spin state Co$^{3+}$ ions are reduced by Ga$^{3+}$ substitution.
We evaluate from the solid line shown in Fig. 5(b) the critical concentration to be $y = 0.25$ at which the magnetic volume fraction disappears.
This suggests that the content of the high-spin state Co$^{3+}$ ion is about 0.25 per formula unit in LaCo$_{0.8}$Rh$_{0.2}$O$_{3}$.

Let us evaluate the temperature dependence of the concentration of the high-spin state Co$^{3+}$ from the temperature dependence of the magnetization shown in Fig. 2(a).
When the number of the magnetic ions \textit{N} in Eq. (1) depends on temperature ($N = N(T)$), Eq. (1) is rewritten as
\begin{equation}
\frac{\mu_{0} H}{M}=\frac{3k_{\rm B}}{N(T)\mu_{\rm B}^{2}\mu^{2}_{\rm eff}}T-\frac{3k_{\rm B}}{N(T)\mu_{\rm B}^{2}\mu^{2}_{\rm eff}}\theta.
\label{reciCWlaw}
\end{equation}
The Curie temperature $\theta$ is proportional to the number of the nearest neighbor sites of the magnetic ions \textit{z} within the mean field approximation.
When the content of high-spin state Co$^{3+}$ is changed with temperature, we can simply assume that \textit{z} is proportional to \textit{N}(\textit{T}).
Since $\theta$ is proportional to \textit{N}(\textit{T}) in this approximation, the second term of Eq. (2) is independent of temperature.
Then we can evaluate the second term of Eq. (2) expressed as
\begin{equation}
A = \frac{3k_{\rm B}}{N(T)\mu_{\rm B}^{2}\mu^{2}_{\rm eff}}\theta (T),
\label{A}
\end{equation}
by extrapolating the reciprocal magnetic susceptibility towards 0 K.
After determining the value of $A$, we obtain $N(T)$ from Eq. (2) as
\begin{equation}
N(T)=\frac{3k_{\rm B}T}{\mu_{\rm B}^{2}\mu^{2}_{\rm eff}(\frac{\mu_{0}H}{M}+A)},
\end{equation}
where we set $\mu^{2}_{\rm eff}$ = 24 as the high-spin-state value.
\begin{figure}[t]
\begin{center}
\includegraphics[width=70.0mm,clip]{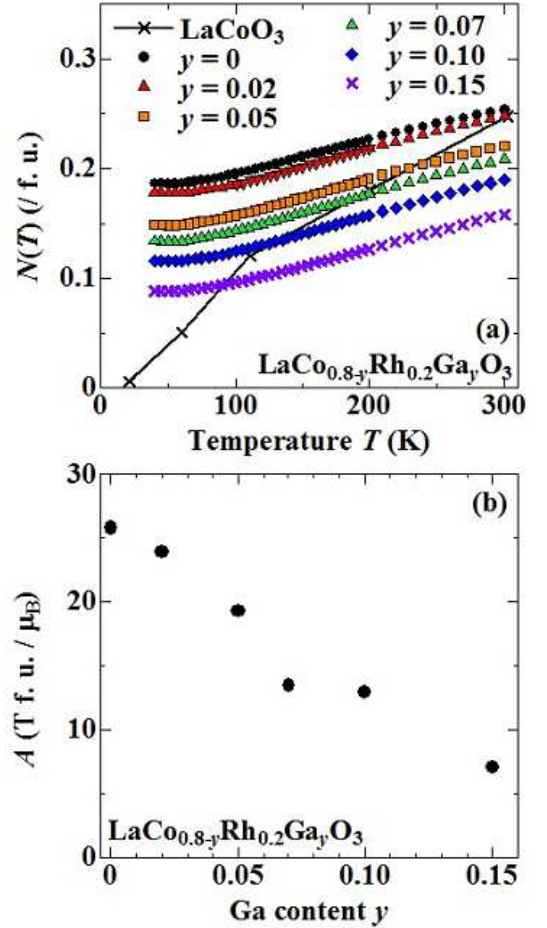}\\
\caption{(Color online) (a) The content of the high-spin state Co$^{3+}$ (\textit{N}(\textit{T})) in LaCo$_{0.8-y}$Rh$_{0.2}$Ga$_{y}$O$_{3}$ ($0 \leq y \leq 0.15$) plotted as a function of temperature.
For comparison, $N(T)$ in LaCoO$_{3}$ is also plotted (taken from Ref. 8).
The detailed calculation method is described in the text. (b) Ga content dependence of the parameter \textit{A} for LaCo$_{0.8-y}$Rh$_{0.2}$Ga$_{y}$O$_{3}$ ($0 \leq y \leq 0.15$) (see text).}
\end{center}
\end{figure}
Figure 6(a) shows the thus obtained $N(T)$ in LaCo$_{0.8-y}$Rh$_{0.2}$Ga$_{y}$O$_{3}$ ($0 \leq y \leq 0.15$), together with $N(T)$ in LaCoO$_{3}$ reported by Haverkort \textit{et al.}~\cite{PRL.97.176405}
Note that \textit{N}(300 K) are almost the same value between LaCo$_{0.8}$Rh$_{0.2}$O$_{3}$ and LaCoO$_{3}$, which indicates that the Rh$^{3+}$ substitution does not change the content of the high-spin state Co$^{3+}$.~\cite{comparison}
With decreasing temperature, $N(T)$ of LaCoO$_{3}$ decreases to zero at low temperature.
On the contrary, $N(T)$ of LaCo$_{0.8}$Rh$_{0.2}$O$_{3}$ is 0.2 per formula unit at 40 K, which indicates that the Rh$^{3+}$ substitution stabilizes the high-spin state Co$^{3+}$ down to low temperatures.
We further emphasize that \textit{N}(40 K) of LaCo$_{0.8}$Rh$_{0.2}$O$_{3}$ is close to the critical concentration of 0.25 evaluated in Fig. 5(b).
$N(T)$ of LaCo$_{0.8-y}$Rh$_{0.2}$Ga$_{y}$O$_{3}$ ($0 \leq y \leq 0.15$) exhibits a downward parallel shift, which suggests that the Ga$^{3+}$ substitution effect is static and local.
Figure 5(b) shows the Ga content dependence of the constant \textit{A}, which linearly decreases with increasing Ga content.

Finally, we discuss an origin of Rh$^{3+}$ and Ga$^{3+}$ substitution effect for Co$^{3+}$ in LaCoO$_{3}$.
Our experimental results indicate that Rh$^{3+}$ and Ga$^{3+}$ seem to replace the low-spin and high-spin state Co$^{3+}$, respectively.
In HS - LS model, it is proposed that the high-spin state Co$^{3+}$ stabilizes the nearest neighbor low-spin state Co$^{3+}$ and that the low-spin state Co$^{3+}$ has opposite interaction.~\cite{PR.155.932}
We extend the idea, and propose that Rh$^{3+}$ and Ga$^{3+}$ stabilize the neighboring high-spin state and low-spin state Co$^{3+}$, respectively. 
A similar behavior is observed in La$_{1-x}$Sr$_{x}$CoO$_{3}$ ($x < 0.01$), where one low-spin state Co$^{4+}$ ion and six intermediate-spin state Co$^{3+}$ ions form a heptamer polaron (spin-state polaron) at low temperature.~\cite{PhysRevB.53.R2926, PRL.101.247603}
We expect that Rh$^{3+}$ and Ga$^{3+}$ substitutions also induce another type of spin-state polaron.
Further microscopic measurements are needed in order to clarify this phenomenon, which are in progress.
\section{Summary}
We have measured the magnetization and the lattice volume for LaCo$_{0.8-y}$Rh$_{0.2}$Ga$_{y}$O$_{3}$ ($0 \leq y \leq 0.15$).
The magnetization is reduced by Ga$^{3+}$ substitution more strongly than by Rh$^{3+}$ substitution, and the lattice volume is almost independent of Ga$^{3+}$ substitution.
These results suggest that a Ga$^{3+}$ ion is selectively substituted for a high-spin state Co$^{3+}$.
We find that one Ga$^{3+}$ ion removes an effective magnetic moment of 4.6 $\mu_{\rm B}$, which is close to the effective magnetic moment (4.9 $\mu_{\rm B}$) of high-spin state Co$^{3+}$.
We further find that 25 \% of Co$^{3+}$ ions survives as the high-spin state for LaCo$_{0.8}$Rh$_{0.2}$O$_{3}$ at 2 K. 
We evaluate the temperature dependence of the concentration of the high-spin state Co$^{3+}$ in LaCo$_{0.8-y}$Rh$_{0.2}$Ga$_{y}$O$_{3}$ ($0 \leq y \leq 0.15$), and find that the Rh$^{3+}$ substitution partially stabilizes the high-spin state of Co$^{3+}$ to the lowest temperature.
We conclude that Ga$^{3+}$ and Rh$^{3+}$ act as if they were substituted for high-spin state Co$^{3+}$ and low-spin state Co$^{3+}$, respectively.
These results strongly suggest that the magnetically excited state of LaCoO$_{3}$ is a mixed state of the high-spin and the low-spin states Co$^{3+}$.

We would like to thank T. Sudayama, H. Nakao, T. Fujita, and M. Hagiwara for fruitful discussion.
This work is partially supported by Program for Leading Graduate Schools, Japan Society for Promotion of Science and by Advanced Low Carbon Research and Development Program, Japan Science and Technology Agency.


\begin{thebibliography}{99}
\bibitem{comparison}
The Curie constant in our previous study was evaluated to be 1.3 times larger than the present value, owing to the different analysis.
We do not stick to the absolute value of $N(T)$, but try to show the relative change with substitution.
\end{thebibliography}
\end{document}